\documentclass[preprint]{aastex631}

\usepackage{amsmath}
\usepackage{soul}
\usepackage{enumitem}

\begin{document}

\title{Advancing solar magnetic field extrapolations through multi-height magnetic field measurements}

\author{Robert Jarolim}
\affiliation{University of Graz, Institute of Physics, Universitätsplatz 5, 8010 Graz, Austria}

\author{Benoit Tremblay}
\affiliation{High Altitude Observatory, 3080 Center Green Dr., Boulder, CO 80301, USA}

\author{Matthias Rempel}
\affiliation{High Altitude Observatory, 3080 Center Green Dr., Boulder, CO 80301, USA}

\author{Momchil Molnar}
\affiliation{High Altitude Observatory, 3080 Center Green Dr., Boulder, CO 80301, USA}

\author{Astrid M. Veronig}
\affiliation{University of Graz, Institute of Physics, Universitätsplatz 5, 8010 Graz, Austria}
\affiliation{University of Graz, Kanzelhöhe Observatory for Solar and Environmental Research, Kanzelhöhe 19, 9521 Treffen am Ossiacher See, Austria}

\author{Julia K. Thalmann}
\affiliation{University of Graz, Institute of Physics, Universitätsplatz 5, 8010 Graz, Austria}

\author{Tatiana Podladchikova}
\affiliation{Skolkovo Institute of Science and Technology, Bolshoy Boulevard 30, bld. 1, Moscow 121205, Russia}



\begin{abstract}

Non-linear force-free extrapolations are a common approach to estimate the 3D topology of coronal magnetic fields based on photospheric vector magnetograms. The force-free assumption is a valid approximation at coronal heights, but for the dense plasma conditions in the lower atmosphere, this assumption is not satisfied. In this study, we utilize multi-height magnetic field measurements in combination with physics-informed neural networks to advance solar magnetic field extrapolations. We include a flexible height-mapping, which allows us to account for the different formation heights of the observed magnetic field measurements. The comparison to analytical and simulated magnetic fields demonstrates that including chromospheric magnetic field measurements leads to a significant improvement of our magnetic field extrapolations. We also apply our method to chromospheric line-of-sight magnetograms, from the Vector Spectromagnetograph (VSM) on the Synoptic Optical Long-term Investigations of the Sun (SOLIS) observatory, in combination with photospheric vector magnetograms, from the Helioseismic Magnetic Imager (HMI) onboard the Solar Dynamic Observatory (SDO). The comparison to observations in extreme ultraviolet wavelengths shows that the additional chromospheric information leads to a better agreement with the observed coronal structures. In addition, our method intrinsically provides an estimate of the corrugation of the observed magnetograms. With this new approach, we make efficient use of multi-height magnetic field measurements and advance the realism of coronal magnetic field simulations.

\end{abstract}



\section{Introduction} \label{sec:intro}

The magnetic field of the Sun is primarily observed from spectral lines that originate low in the solar atmosphere. To obtain a better understanding of the build-up and release of magnetic energy in the solar atmosphere, a full 3D understanding of the magnetic field topology is required \citep{2018SSRv..214...46G, 2017SSRv..210..249W}. A frequently applied approach are non-linear force-free (NLFF) magnetic field extrapolations, where the photospheric vector magnetogram is used as boundary condition and plasma density is assumed to be negligible \citep{wiegelmann2021sffmf}. This requires solving a system of coupled partial differential equations in the form of the divergence-free equation
\begin{equation}
    \vec{\nabla} \cdot \vec{B} = 0 \,,
    \label{eq:div}
\end{equation}
and the force-free equation
\begin{equation}
    \vec{J} \times \vec{B} = 0\,,
    \label{eq:ff}
\end{equation}
where $\vec{B}$ refers to the magnetic field vector and $\vec{J}$ to the electric current density. The force-free model builds on the assumption that the plasma density is low, and therefore the ratio of plasma pressure and magnetic pressure is small (plasma beta $\beta_{\rm plasma} \ll 1$), which is not satisfied in the solar photosphere but which can be validly assumed for heights of $\sim$400\,km and above \citep{1995ApJ...439..474M}. Commonly relying on regularly performed photospheric vector magnetic field measurements as an input, extrapolations therefore deal with an inherently ill-posed problem, where a certain adjustment of the photospheric vector field prior to the supply as an input to the extrapolation or during relaxation were shown to obtain solutions of improved quality \citep[see][]{wiegelmann2006preprocessing,wiegelmann2010methood}.

In \cite{jarolim2023nf2} a physics-informed neural network (PINN) was used to intrinsically find a trade-off between the aforementioned physical model constraints and the observed magnetogram, where the obtained NLFF models adhere to a similar degree of quality and realism. This property is enabled by the ability of PINNs to smoothly integrate noisy data and incomplete physical models \citep{raissi2019pinns, karniadakis2021physicsml}.

In this study, we propose a novel approach based on PINNs that in addition incorporates also magnetic field measurements at multiple optical depths $\tau$ (e.g., chromospheric measurements), in order to achieve more realistic coronal magnetic field extrapolations. Here, we assume that the force-free assumption is applicable in the solar chromosphere and that the additional observational constraint, in the form of an internal condition, can achieve a more realistic field estimate of the solar corona.
For observations from upper atmospheric layers (lower optical depth $\tau$), we expect a stronger variation in the height of formation of the line. Consequently, the magnetic field observations do not correspond to a geometrically flat surface of the solar atmosphere, but are corrugated surfaces, that can span multiple Mm \citep{carson1976stellar_opacity, quintero2023atmospheric_information}.
To account for this, we include in our study a height mapping module that allows our neural network to dynamically adjust the estimated local (i.e., pixel-wise) geometrical height of the observed magnetic field to be consistent with the extrapolated field. In the same way, our method adjusts the extrapolated field to be compatible with the observations. In addition to the improved magnetic field extrapolation, this approach allows us to extract also the geometrical height variation of the observed magnetic field from the model.

Magnetic field extrapolations are typically performed using photospheric magnetograms \citep[e.g.,][]{wiegelmann2010methood, wheatland2011nlff}. In \cite{yelles2012chromospheric_extrapolation}, chromospheric slices were used for magnetic field extrapolations. In \cite{vissers2022sst_multi_height}, the similarity between the chromospheric magnetic field, inferred from observations, and the field obtained from a magnetohydrostatic extrapolation are compared. In this study, we consistently incorporate both photospheric and chromospheric magnetic field observations in our non-linear force-free model.

We demonstrate the advantage of our approach by applying our method to a semi-analytical magnetic field (Sect. \ref{sec:analytic}), and to a realistic solar magnetic field simulation (MURaM; Sect. \ref{sec:muram}). Both test cases provide a ground-truth reference to evaluate the performance of our method. In Sect. \ref{sec:observation}, we apply our method to chromospheric observations from SOLIS/VSM in combination with photospheric magnetograms from SDO/HMI.

\section{Method}
\label{sec:method}

\begin{figure*}
    \centering
    \includegraphics[width=\linewidth]{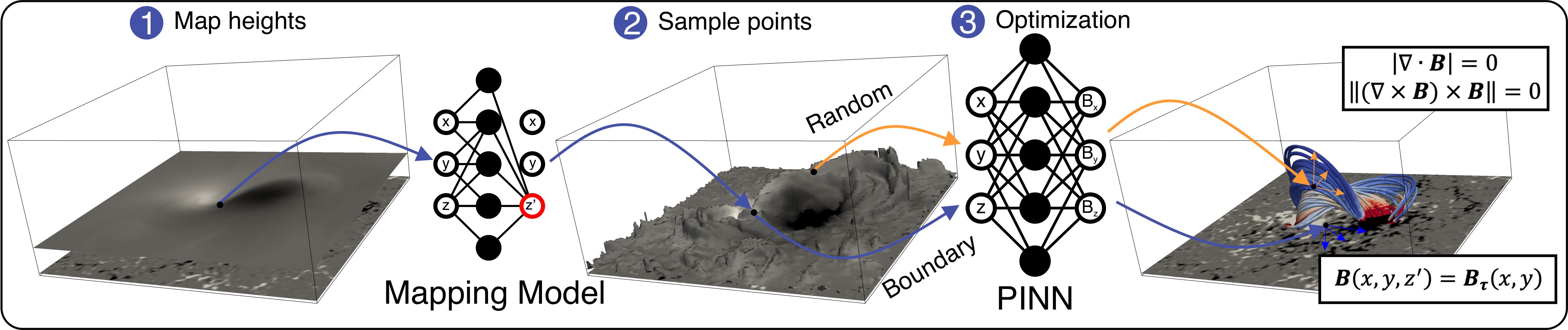}
    \caption{Overview of our proposed method. We extend the method presented in \cite{jarolim2023nf2} by including a height-mapping model. The height-mapping model takes coordinate points ($x, y, z$) and adjusts the the vertical component to $z'$. From the corrugated surface we sample both boundary values and random coordinates for our model training. The PINN and the height-mapping model, are updated simultaneously by optimizing for the force-free condition and matching the boundary condition of the observed magnetograms.}
    \label{fig:method}
\end{figure*}

In this study, we follow the approach from \cite{jarolim2023nf2} for coronal magnetic field extrapolations. We use a neural network as a mesh-free representation of the simulation volume, mapping coordinate points ($x, y, z$) to the respective magnetic field vector ($B_x, B_y, B_z$). Since neural networks are fully-differentiable, we can compute smooth derivatives of the output magnetic field vectors with respect to the input coordinates (e.g., $\partial B_x/ \partial x, \partial B_x / \partial y$). These are then used to construct the divergence-free and force-free equations.
This provides the physical loss terms for our model optimization, starting with the force-free loss
\begin{equation}
\label{eq:l_ff}
    L_{\rm ff} = \frac{\lVert(\vec{\nabla} \times \vec{B}) \times \vec{B}\rVert^2}{\lVert \vec{B} \rVert^2 + \epsilon},
\end{equation}
and the divergence-free loss
\begin{equation}
\label{eq:l_div}
    L_{\rm div} = |\vec{\nabla} \cdot \vec{B}|^2.
\end{equation}
Here, $\vec{B} \equiv \vec{B}(x,y,z)$ refers to the neural network representation of the magnetic field vector at a given point  $(x, y, z)$ and $\epsilon=10^{-6}$ is added for numerical stability.

In addition, we optimize our neural network to reconstruct a magnetic field $\vec{B}$ that matches the boundary and internal conditions
$\vec{B}_{\tau}(x, y)$, where $\vec{B}_{\tau}$ refers to an observed magnetogram at the optical depth $\tau$. We optimize for this condition by sampling coordinates from the vector magnetogram and minimizing the
component-wise distance between the observed $\vec{B}_\tau$ and modeled $\vec{B}$ magnetogram at the same position.
The loss term is then given by
\begin{equation}
    \label{eq:l_b}
    L_{\rm  B} = \lVert \vec{B} - \vec{B}_{\tau} \rVert^2.
\end{equation}

In this study, we add multiple height surfaces to our extrapolation. We account for this by adding a height-mapping module that maps coordinate points $(x, y, z)$ to a new height surface $(x, y, z')$ (only the vertical coordinate is updated). Here, $z$ is a constant average geometrical height, estimated for each $\tau$ surface and remapped in dependence of the spatial location $z'(x, y, z)$. The resulting mapped coordinate points and randomly sampled points are then used as input to the same PINN as for the regular extrapolation (see Fig. \ref{fig:method}). This enables us to intrinsically model the corrugation of the observed magnetograms at constant optical depths $\tau$ as part of the optimization procedure. For our initial coordinates, we estimate the average line-formation height $z$.

For the optimization, the model is iteratively updated by randomly sampling points from the simulation volume and from the boundary and internal conditions. We emphasize that this approach solves a single simulation and does not rely on an underlying ground-truth data set. The application to a new active region requires a re-training of the model.

For our neural networks, we use the SInusoidal REpresentation Networks (SIREN) architecture from \cite{sitzmann2020siren}, analogously to \cite{jarolim2023nf2}. For the height mapping and PINN model, we use 4 and 8 hidden layers respectively. Each layer consists for 256 neurons. For a comparison of different model architectures see \cite{jarolim2023nf2}. For the height-mapping model, we use a positional encoding of the input coordinates \citep{mildenhall2020nerf}. As output, we use a sigmoid function and scale the predicted coordinate to the specified height range (such that $z' = [z'_{\rm min}, z'_{\rm max}]$).

\subsection{Regularization}

We include two additional regularization terms into our model optimization. We compute the height regularization
\begin{equation}
    L_{\text{h}} = \frac{|z' - z|}{\Delta z + \epsilon},
\end{equation}
where $\Delta z$ corresponds to the height range ($z'_{\rm max} - z'_{\rm min}$) and $\epsilon = 10^{-6}$ is added for numerical stability. With this, we prefer mappings that are close to the estimated height.

We introduce a regularization for missing components (e.g., the missing horizontal vector components), by optimizing for minimum energy solutions. We use
\begin{equation}
    L_{\text{NaN}} =  B_{\text{NaN}} ^2,
\end{equation}
where $B_{\text{NaN}}$ refers to the unspecified components (NaN values). With this, we prevent unrealistic strong magnetic fields of undefined field components.

The combined loss is then computed by the weighted sum of the individual terms
\begin{equation}
    L = \lambda_{\text{ff}} L_{\text{ff}} + \lambda_{\text{div}} L_{\text{div}} + \lambda_{\rm B} L_{\rm B} + \lambda_{\text{h}}  L_{\text{h}} + \lambda_{\text{NaN}}  L_{\text{NaN}}.
\end{equation}
For all our experiments we set $\lambda_{\text{ff}} = \lambda_{\text{div}} = 0.1$, and use a scheduled weight for $\lambda_{\rm B}$, where we exponentially decay from $1000$ to $1$ over $10^{5}$ iterations \citep[for a detailed parameter discussion see ][]{jarolim2023nf2}. For our regularization we set $\lambda_{\text{h}} = 10^{-3}$ and $\lambda_{\text{NaN}} = 10^{-2}$. For all our experiments we start from a randomly initialize neural network and train our model iteratively, where we sample $10^{4}$ coordinate points from the boundary condition and $2 \times 10^{4}$ coordinate points randomly and continuously from the simulation volume. The loss per update step is averaged over the full batch. $L_{\rm B}$, $L_{\text{h}}$, and $L_{\rm NaN}$ are computed over the sampled boundary and internal conditions, while $L_{\text{div}}$ and $L_{\text{ff}}$ are computed over all sampled points. For our extrapolations we use $2 \times 10^{5}$ iterations. The analytical magnetic field solutions in Sect. \ref{sec:analytic} are trained for $10^{5}$ iterations. We use the Adam optimizer \citep{kingma2014adam} with beta $\beta_{\rm Adam}$ of (0.9, 0.999), and exponentially schedule the learning rate from $5 \times 10^{-4}$ to $5 \times 10^{-5}$ over the full length of the training.

After model training, we obtain the extrapolated magnetic field by sampling the magnetic field vectors at each grid cell in our simulation volume, using the spatial resolution of the boundary condition. The resulting data cube is then used for further evaluation, where we use finite-differences to compute derivatives of the magnetic field (e.g., divergence, current density).

\subsection{Data}

For the verification of our method, we consider two data sets that provide a ground-truth reference. We use a semi-analytical magnetic field solution to estimate the models ability to extrapolate smooth force-free fields in an ideal setting (Sect. \ref{sec:data_analytical}). We further use a snapshot from a MHD simulation, which does not satisfy the force-free assumption. Therefore, this data set provides a more realistic estimation of the model performance for the application to observational data (Sect. \ref{sec:data_muram}).
For the application to observational data, we use photospheric and chromospheric magnetograms (Sect. \ref{sec:data_observation}).
For the comparison with potential magnetic fields, we compute the solution using the Green's function approach as proposed by \cite{sakurai1982green}. This simpler model assumes a current-free field and serves as additional baseline.

\subsubsection{Semi-analytical field}
\label{sec:data_analytical}

We apply our method to extracted slices from the "Low and Lou" model \citep{lowlou1990analytic}. This magnetic field model is analytical except for the numerical solution of an ordinary differential equation in spherical coordinates with eigenvalues $m$ and $n$. The obtained fields are axisymmetric, but by rotating the symmetry axis by a rotation angle $\Phi$ and placing the singularity at a depth $l$ below the photosphere, we can mimic coronal fields that are useful for assessing the performance of NLFF models for the case of large-scale smooth currents distributed in the entire test volume \citep{wiegelmann2021sffmf, schrijver2006NLFFcomparison, 2000ApJ...540.1150W}. Here we use the Case~I configuration from \cite{schrijver2006NLFFcomparison}, where six different state-of-the-art NLFF methods for 3D coronal field extrapolation have been systematically compared. The configuration uses $n=1$, $m=1$, $l=0.3$ and $\Phi=\pi/4$, in a volume spanning $64^3$ grid points.

\subsubsection{MURaM snapshot}
\label{sec:data_muram}

We use a realistic simulation of the solar magnetic field from the MPS/University of Chicago Radiative MHD code \citep[MURaM: ][]{voegler2005muram, rempel2017muram}. The setup simulates the emergence of a kink-unstable twisted bipole in a domain of $98.304$~Mm horizontal and $82.944$~Mm vertical extent. The photosphere is located about $7.8$~Mm above the bottom boundary. The simulation has a resolution of 0.192 Mm per pixel in the horizontal dimensions and 0.064 Mm per pixel for the vertical dimension. For our application we reduce the resolution of all dimensions to 0.384 Mm per pixel (binning by a factor of 2 and 6 for the horizontal and vertical axis, respectively), similar to the HMI resolution of 0.36 Mm per pixel. The snapshot we use as reference in this investigation is the same as used in \cite{2021ApJ...917...27J} to highlight the chromospheric corrugation and diagnostic potential of several chromspheric lines in the near UV \citep[see, Figure 2 in][]{2021ApJ...917...27J}.

We extract magnetic field layers at Rosseland optical depths $\tau = \{10^{-6}, 10^{-5}, 10^{-4}, 10^{-3}, 10^{-2}, 10^{-1}\}$, by using the optical depth scale of a vertical ray computed with Rosseland mean opacities. We compute the optical depth $\tau$ at each point in the simulation volume $(x, y, z)$ according to
\begin{equation}
    \tau(x, y, z) = \int_{z}^{z_{\rm top}} \kappa(x, y, z) \, \rho(x, y, z) \, dz,
\end{equation}
where $z$ refers to the geometrical height, $z_{\rm top}$ to the top of the simulation volume ($\sim$75 Mm above the photosphere), $\kappa$ to the opacity, and $\rho$ to the plasma density, which are taken from the MURaM simulation.
We select surfaces of constant $\tau$, where we compute the geometrical height $z$ and select the magnetic field vector at the corresponding points. The extracted height surfaces serve as reference for the evaluation of the height mapping module, while the magnetic field is used as input to our model. The optical depth range corresponds to the height from the photosphere to upper chromosphere. Note that the extracted magnetic fields correspond to a thin layer, resembling optically-thick spectral lines. Since this simulation describes active flux emergence, the deformation of the higher chromospheric $\tau$ surfaces is considerable and exceeds $10$~Mm for $\tau = 10^{-6}$. This example was specifically chosen to estimate the model performance for a broad range of corrugations. The extracted $\tau$ surfaces and magnetic fields are shown in Fig. \ref{fig:tau}.

\subsubsection{Observations}
\label{sec:data_observation}

To test the method on real solar observations, we use cotemporal
synoptic chromospheric line-of-sight magnetic fields from the Vector Spectromagnetograph
(VSM) on the Synoptic Optical Long-term Investigations of
the Sun \citep[SOLIS: ][]{2003SPIE.4853..194K} and
photospheric magnetic vector data from the Helioseismic
Magnetic Imager \citep[HMI: ][]{schou2012hmi} onboard the Solar Dynamic
Observatory \citep[SDO; ][]{pesnell2012sdo} spacecraft. SOLIS/VSM provides full-disk
chromospheric magnetic field estimates
inferred from the polarization signatures of the chromospheric
\ion{Ca}{2} 854.2 nm line. In particular,
SOLIS/VSM provides line-of-sight magnetic
field strength calculated with the weak field approximation applied to
the chromospheric core of the \ion{Ca}{2} 854.2 nm line.
We note that VSM also provides photospheric data, but the corresponding photospheric magnetogram is recorded with a delay of one hour. Therefore, we use a combination of chromospheric VSM and photospheric HMI data, to ensure the best temporal alignment.

We chose observations of NOAA AR 11166 on from 2011 March 9 as the example case
for this study,  due to its proximity to disc center and
the favorable observing conditions at the
SOLIS observing site. We use the SOLIS/VSM \ion{Ca}{2} 854.2 nm dataset
taken at 16:33 UT and the corresponding 16:36 UT dataset from the
SHARPs~\citep{bobra2014sharps} AR HMI catalogue.
We reproject the helioprojective SOLIS/VSM data
to the CEA projection of the SHARPs dataset using the
SunPy library~\citep{sunpycommunity2020}.


\subsection{Analysis Tools}

\subsubsection{Metrics}

For the quantitative evaluation we use the metrics from \cite{schrijver2006NLFFcomparison}. We denote the magnetic field vectors at point $i$ as $\vec{B}_{\text{ref},i}$ and $\vec{B}_{\text{ext},i}$ for the reference solution and the extrapolated magnetic field respectively. The vector correlation coefficient ($C_{vec}$) compares the local characteristics of the magnetic field vectors
\begin{equation}
    C_{vec}(\vec{B}_{\rm ref}, \vec{B}_{\rm ext}) = \frac{\sum_i \vec{B}_{\text{ref}, i} \cdot \vec{B}_{\text{ext},i}}{\left(\sum_i \lVert \vec{B}_{\text{ref}, i} \rVert^2 \sum_i \lVert \vec{B}_{\text{ext},i} \rVert^2\right)^{1/2}} \,.
\end{equation}
The metric gives 0 for no correlation (perpendicular vectors) and 1 for identical vectors.
The second metric measures the angle between the magnetic field solutions based on the Cauchy-Schwarz inequality (cosine similarity index)
\begin{equation}
    C_{CS}(\vec{B}_{\rm ref}, \vec{B}_{\rm ext}) = \frac{1}{M} \sum_i \frac{\vec{B}_{\text{ref}, i} \cdot \vec{B}_{\text{ext},i}}{\lVert \vec{B}_{\text{ref}, i} \rVert \lVert \vec{B}_{\text{ext},i} \rVert} \,,
\end{equation}
where $M$ refers to the total number of grid points in the volume.
$C_{CS}$ ranges from -1 (anti-parallel vectors) to 1 (parallel vectors), with a value of 0 indicating perpendicular vectors.
To measure the difference between vectors, we compute the mean error normalized by the average vector norm
\begin{equation}
    E_n (\vec{B}_{\rm ref}, \vec{B}_{\rm ext}) = \frac{\sum_i \lVert \vec{B}_{\text{ext},i} - \vec{B}_{\text{ref}, i} \rVert}{\sum_i \lVert \vec{B}_{\text{ref}, i} \rVert} \,,
\end{equation}
and the mean error normalized per vector
\begin{equation}
    E_m (\vec{B}_{\rm ref}, \vec{B}_{\rm ext}) = \frac{1}{M} \sum_i \frac{\lVert \vec{B}_{\text{ext},i} - \vec{B}_{\text{ref}, i} \rVert}{\lVert \vec{B}_{\text{ref}, i} \rVert} \,.
\end{equation}
Here, the best performance is achieved when $E_m = E_n = 0$. We set $E'_{m,n} = 1 - E_{m,n}$, such that the best attainable performance corresponds to 1, to allow for an easier comparison with the previous metrics.

We further compare the global magnetic field, where we compute the modeled magnetic energy relative to the reference magnetic energy

\begin{equation}
    \varepsilon(\vec{B}_{\rm ref}, \vec{B}_{\rm ext}) = \frac{\sum_i \lVert \vec{B}_{\text{ext},i} \rVert ^ 2}{\sum_i \lVert \vec{B}_{\text{ref}, i} \rVert ^ 2} \,,
\end{equation}
where the best performance equals to 1. Values $\varepsilon >1$ and $\varepsilon < 1$ correspond to an over- or under-estimation, respectively.

To quantify the divergence-freeness of the magnetic field $\vec{B}$, we compute the normalized divergence
\begin{equation}
    L_{div,n}(\vec{B}) = \sum_i |\vec{\nabla} \cdot \vec{B}_{i}| / \lVert \vec{B}_{i} \rVert \,.
\end{equation}
To quantify the force-freeness of the magnetic field, we use the current weighted average sine of the the angle between the magnetic field and its current density vector \citep[c.f.,][]{schrijver2006NLFFcomparison}:
\begin{equation}
    \sigma_J(\vec{B}) = \left( \sum_i \frac{\lVert \vec{J}_i \times \vec{B}_{i} \rVert}{\lVert \vec{B}_{i} \rVert} \right) / \sum_i \lVert \vec{J}_i \rVert \,.
\end{equation}

\subsubsection{Magnetic mapping and twist}

In order to characterize the differences in magnetic connectivity in the NLFF modeling based solely on photospheric magnetic field information and including in addition magnetic information at a chromospheric level, we employ two measures that are based on magnetic field lines traced within the NLFF models, via the application of the fourth-order Runge-Kutta method described in \citep{2016ApJ...818..148L}.

First, in order to pin down regions of strong connectivity gradients, we compute the squashing factor $Q$. Therefore, we characterize the mapping of elementary flux tubes (approximated by point-wise traced magnetic field lines in the NLFF models), using the Jacobian matrix
\begin{equation}
    D_{12} = \left[ \frac{\partial\vec{r}_2}{\partial\vec{r}_1} \right] =
    \left(
    \begin{matrix}
    \partial x_2/\partial x_1 & \partial x_2/\partial y_1 \\
    \partial y_2/\partial x_1 & \partial y_2/\partial y_1
    \end{matrix}
    \right) \equiv
    \left(
    \begin{matrix}
    a & b \\
    c & d
    \end{matrix}
    \right),
\end{equation}
associated to the mapping of the two footpoints of a field line (elementary flux tube) $\prod_{12}: \vec{r}_1(x_1,y_1) \mapsto \vec{r}_2(x_2,y_2)$. Then, following \cite{2002JGRA..107.1164T}, the squashing factor is defined as
\begin{equation}
    Q = \frac{a^2+b^2+c^2+d^2}{|B_{n,1}(y_1,y_1)/B_{n,2}(x_2,y_2)|},
\end{equation}
where $B_{n,1}(y_1,y_1)$ and $B_{n,2}(x_2,y_2)$ are the magnetic field components normal to the plane of the footpoints (the lower boundary in our case), and their ratio is equivalent to the determinant of $D_{12}$.

Second, in order to indicate how many turns two (presuming infinitesimally closed) field lines wind about each other, we compute the twist number \citep{2006JPhA...39.8321B}
\begin{equation}
    T_w = \int_L \frac{(\vec{\nabla} \times \vec{B}) \cdot \vec{B}}{4\pi B^2} ~dl \,,
\end{equation}
where the integrand can be regarded as a local twist density along of a field line. If $\vec{\nabla} \times \vec{B} = \alpha \vec{B}$, with $\alpha$ being the force-free parameter (constant along of each individual field line in the case of a perfectly force-free field), then $T_w = \frac{1}{4 \pi}\int_L \alpha \,dl$. The twist number distribution in a vertical plane in the model volume is achieved by assigning the twist number of each field line to the position where this field line threads the plane.

\section{Results}

To assess the validity of our approach, we consider three evaluation schemes. (1) We use a semi-analytical magnetic field solution $\vec{B}_{\rm ref}$ from which we extract geometrical slices $z_{\rm ref}$ and generate artificially corrugated slices $\tau_{\rm ref}$ (to mimic constant $\tau$ surfaces). Here we consider the ideal case of a force-free magnetic field (Sect. \ref{sec:analytic}). (2) We extract reference magnetic fields $\vec{B}_{\rm ref}$ at constant $\tau_{\rm ref}$ from a MURaM simulation snapshot. In this case, we are dealing with a realistic magnetic field that is not force-free and where the corrugation is computed based on the simulated optical depth $\tau$ (Sect. \ref{sec:muram}). (3) We utilize observational data of AR 11166 (SHARP 401) from SOLIS in conjunction with SDO/HMI vector magnetograms to estimate both the topology and height of formation of the magnetic field. We compare our magnetic field extrapolation $\vec{B}_{\rm ext}$ to EUV observations in order to assess the realism of the extrapolated magnetic field topology (Sect. \ref{sec:observation}).

\subsection{Evaluation with analytical fields}
\label{sec:analytic}

We use the semi-analytical field $\vec{B}_{\rm ref}$ described in Sect. \ref{sec:data_analytical} and consider four principal configurations for our extrapolations (Table \ref{table:analytic}). For all configurations we normalize the magnetic field strength to 300 Gauss and the spatial scaling to 64 pixels.
First, we perform a standard extrapolation that uses only the bottom boundary as an input. The upper boundaries are either taken from a potential-field solution or from the semi-analytical field. Note that the boundary and internal conditions are not strictly enforced and are dynamically adjusted throughout the model training. For the remaining configurations we impose no side-boundaries. Therefore, $L_{\rm B}$ is computed based on the bottom and internal boundary conditions (Eq. \ref{eq:l_b}), while the remaining volume, including the lateral and top boundaries, is determined by minimizing the force-free condition at randomly sampled points within the simulations volume (Eqs. \ref{eq:l_ff} and \ref{eq:l_div}). In other words, the lateral boundaries are not fixed during training but are solely determined by the physical equations.
For the second configuration, we extract geometrical slices at the boundary and within the simulation volume which spans a height of 64 pixels. We extract geometrical slices at heights $z_{\rm ref} = \{0, 8, 16, 32\}$ pixels and compare extrapolations $\vec{B}_{\rm ext}$ that use a combination of two to four slices.
For the second configuration, we extract geometrical slices from the simulation volume, which spans 64 pixels in height, starting at the lower boundary $z_{\rm ref} = 0$ pixels and at increasing heights within the volume  $z_{\rm ref} = \{8, 16, 32\}$ pixels.
We compare extrapolations $\vec{B}_{\rm ext}$ that use a combination of the boundary condition and one to three internal conditions.

For the third configuration, we use artificial $\tau$ surfaces instead of constant geometrical height surfaces. More specifically, we generate corrugated magnetograms by extracting magnetic field values at locations $(x, y, z_{\rm ref})$ where $z_{\rm ref}$ is taken from a 2D normal distribution
\begin{equation}
    z_{\rm ref}(x, y, h) = h \, \, \exp\left(-\frac{(x^2+y^2)}{2 \sigma(h)^2}\right) \,,
\end{equation}
where $x$ and $y$ ranges between $[-32, 32]$ pixels (i.e., the horizontal limits of the volume), $h$ is the constant geometrical height that ranges between $[0, 64]$ pixels, and the width $\sigma$ varies according to
\begin{equation}
    \sigma(h) = \frac{h}{64} + 0.5 \,.
\end{equation}
We extract corrugated magnetograms at heights $h = \{0, 8, 16, 32\}$ pixels. For the slice at $h=0$, we use no corrugation (i.e., we use a geometrical slice as the lower boundary). For the input of our height mapping module, we set the initial heights $z$ to half of the maximum and the range between 0 and the maximum value. We found that for this ideal setting, height regularization is not necessary and we set $\lambda_{h} = 0$.

For the fourth configuration, we utilize the $\tau$ surfaces from the third configuration, but only extract the vertical component of the magnetic field ($B_{{\rm ref}, z}$) for the internal conditions. In other words, we use the full vector information at the lower boundary and only the vertical component for the corrugated surfaces.

Table \ref{table:analytic} summarizes our results. For the first configuration, it can be seen that we achieve a nearly identical magnetic field solution for the extrapolation that uses the full boundary conditions. In contrast, the usage of a potential field approximation leads to a lower performance than the open configuration. This suggest that a wrong upper-boundary condition can strongly affect the result. We note that we consider here a magnetic field that is largely dominated by the lower-boundary condition, while a potential field approximation becomes more important when dealing with larger domains that deviate from the force-free assumption (see e.g. Sect. \ref{sec:muram}).

The second configuration reveals that additional information from geometrical slices leads to a constant improvement over the open configuration, achieving the best results with a maximum number of four slices, that even exceed the configuration where all side-boundaries are specified (full). From the comparison of $\{0, 16\}$ and $\{0, 8, 16\}$, as well as $\{0, 32\}$ and $\{0, 8, 16, 32\}$, we can see that introducing intermediate layers provides only a marginal or no improvement. On the other hand, introducing slices at greater heights (e.g., 32), largely improves the extrapolations. From the analysis of the corrugated surfaces in the third configuration, it can be seen that our height mapping model can properly frame the magnetic field measurements into the extrapolation, where we only note a marginal quality drop. A notable performance decrease occurs when we only consider the vertical component of the corrugated surfaces (fourth configuration). Here, the resulting extrapolation performs throughout worse than the equivalent height surfaces that use the full vector magnetograms. However, the additional height information still provides an improvement over the single-height extrapolation.

The comparison of the height differences ($\Delta h = |z'(x, y, z) - z_{\rm ref}(x, y, h)|$), shows that our mapping model deviates from the ground-truth corrugation by about $1\%$ and between $1-2\%$ for the case of missing horizontal components.

\begin{table*}[!h]
\caption{Quantitative comparison between the semi-analytic magnetic field solutions $\vec{B}_{\rm ref}$ from \cite{lowlou1990analytic} and extrapolations $\vec{B}_{\rm ext}$ performed using different boundary conditions. (1-3) Full: use of all side boundaries for training our model. Potential: upper-boundaries are approximated by a potential field. Open: only the bottom boundary condition is used for model training. (4-8): Geometrical slices at heights $z_{\rm ref} = \{0, 8, 16, 32\}$ pixels. (9-11): Artificially corrugated surfaces $\tau_{\rm ref}$ with maximum geometrical heights of $\{0, 8, 16, 32\}$ pixels. (12-14): Same as for 9-11, but using only the vertical magnetic field component for the upper $\tau$ surfaces. We compare the vector correlation ($C_{vec}$), angular difference ($C_{CS}$), the complement of the normalized vector error ($E'_n= 1 - E_n$), the complement of the mean vector error ($E'_m = 1 - E_m$), relative total magnetic energy ($\varepsilon$), the current-weighted average of the sine of the angle between the magnetic field and electric current density ($\sigma_J$), the normalized divergence $L_{n,div}$ in pixels$^{-1}$, and the distance $\Delta h$ (in pixels and in percentage) between the reference heights $z_{\rm ref}$ and the inferred heights $z'$.}
\label{table:analytic}      
\centering                          
\begin{tabular}{| l || c c c c c c c l l |}        
\hline
 Configuration & $C_{vec}$ & $C_{cs}$ & $E'_n$ & $E'_m$ & $\varepsilon$ & $\sigma_J \times 10^2$ & $L_{div,n} \times 10^2$ & $\Delta$h [pixels] & $\Delta$h [\%] \\    
\hline                        
  \hline
  1 Full & 1.00 & 1.00 & 1.00 & 0.99 & 1.00 & 2.01 & 0.05 & - & - \\
  \hline
  2 Potential & 1.00 & 0.80 & 0.82 & 0.50 & 0.99 & 3.61 & 0.26 & - & - \\
  \hline
  3 Open & 1.00 & 0.92 & 0.95 & 0.71 & 1.47 & 2.30 & 0.06 & - & - \\
  \hline
  \hline
  4 $z_{\rm ref} = \{0, 8\}$ & 1.00 & 0.99 & 0.99 & 0.93 & 1.00 & 2.13 & 0.06 & - & - \\
  \hline
  5 $z_{\rm ref} =\{0, 16\}$ & 1.00 & 0.97 & 0.98 & 0.86 & 1.00 & 2.19 & 0.11 & - & - \\
  \hline
  6 $z_{\rm ref} =\{0, 8, 16\}$ & 1.00 & 0.98 & 0.98 & 0.88 & 1.00 & 2.08 & 0.07 & - & - \\
  \hline
  7 $z_{\rm ref} =\{0, 32\}$ & 1.00 & 1.00 & 0.99 & 0.98 & 1.00 & 2.05 & 0.07 & - & - \\
  \hline
  8 $z_{\rm ref} =\{0, 8, 16, 32\}$ & 1.00 & 1.00 & 1.00 & 0.98 & 1.00 & 2.13 & 0.08 & - & - \\
  \hline
  \hline
  9 $\tau_{\rm ref}(h = \{0, 8\})$ & 1.00 & 1.00 & 0.99 & 0.96 & 1.00 & 2.00 & 0.05 & \{0.07\} & \{0.9\} \\
  \hline
  10 $\tau_{\rm ref}(h  = \{0, 8, 16\})$ & 1.00 & 0.99 & 0.98 & 0.89 & 1.00 & 2.12 & 0.06 & \{0.07, 0.15\} & \{0.9, 0.9\} \\
  \hline
  11 $\tau_{\rm ref}(h  = \{0, 8, 16, 32\})$ & 1.00 & 1.00 & 1.00 & 0.98 & 1.00 & 2.02 & 0.05 & \{0.07, 0.15, 0.36\} & \{0.8, 0.9, 1.1\} \\
  \hline
  \hline
  12 $\tau_{\rm ref}(h  = \{0, 8_z\})$ & 1.00 & 0.98 & 0.97 & 0.86 & 1.00 & 1.95 & 0.05 & \{0.18\} & \{2.3\}\\
  \hline
  13 $\tau_{\rm ref}(h  = \{0, 8_z, 16_z\})$ & 1.00 & 0.98 & 0.97 & 0.85 & 1.00 & 2.16 & 0.07 & \{0.18, 0.33\} & \{2.3, 2.1\} \\
  \hline
  14 $\tau_{\rm ref}(h  = \{0, 8_z, 16_z, 32_z\})$ & 1.00 & 1.00 & 0.99 & 0.94 & 1.00 & 1.95 & 0.05 & \{0.12, 0.25, 0.54\} & \{1.5, 1.6, 1.7\} \\
\hline                                   
\end{tabular}
\end{table*}

\subsection{Evaluation with MURaM $\tau$ surfaces}
\label{sec:muram}

We use the MURaM snapshot introduced in Sect. \ref{sec:data_muram} to evaluate our model in the realistic setting of a non force-free magnetic field computed by the MURaM simulation. For this, we reduce the data to a resolution of 0.384 Mm per pixel and consider the range from 20 to 90 pixels (7.68 to 34.56 Mm; photosphere to coronal heights) for our evaluation. We normalize the magnetic field strength to 2500 Gauss and the spatial scaling to 320 pixels. We extract vector magnetograms $\vec{B}_{\rm ref}$ at optical depths $\tau_{\rm ref} = \{10^{-6}, 10^{-5}, 10^{-4}, 10^{-3}, 10^{-2}, 10^{-1}\}$. The vertical component $B_{\rm ref, z}$ and the corresponding geometrical height of formation of the $\tau_{\rm ref}$ surfaces are shown in Fig. \ref{fig:tau}.

All extrapolations use $\vec{B}_{\rm ref}(\tau = 10^{-1})$ as a lower boundary.
As baseline for our evaluation, we compute three single-height extrapolations.
\begin{itemize}
    \item Potential: A potential field extrapolation;
    \item Extrapolation: A photospheric NLFF extrapolation with open side and top boundaries;
    \item Extrapolation - PB: A photospheric NLFF extrapolation that uses a potential field as side and top boundaries
\end{itemize}
For the multi-height extrapolations we compare four configurations.
\begin{itemize}
    \item Fixed heights: A fixed-height configuration that does not rely on the height model. Vector magnetograms are introduced at the average geometrical heights of formation of $\tau = \{10^{-6}, 10^{-5}, 10^{-4}, 10^{-3}, 10^{-2}, 10^{-1}\}$.
    \item Mapped heights: An ideal setting that uses all available magnetograms in conjunction with the height model, i.e. $\vec{B}_{\rm ref}(\tau = \{10^{-6}, 10^{-5}, 10^{-4}, 10^{-3}, 10^{-2}, 10^{-1}\})$. We allow for the height ranges from the reference height map and set the initial height estimate to the average height.
    \item Realistic vector: A more realistic setting that only incorporates a single chromospheric vector magnetogram as an internal condition, i.e. $\vec{B}_{\rm ref}(\tau = \{10^{-5}, 10^{-1}\})$.
    \item Realistic split: Inversions of spectro-polarimetric measurements can be from different heights for the vertical and horizontal magnetic field components. For this configuration we use the vertical component for the $\tau = 10^{-5}$ layer and the horizontal component of the $\tau = 10^{-4}$ layer, i.e. $\vec{B}_{\rm ref}(\tau = 10^{-1})$, $B_{\rm ref, z}(\tau = 10^{-5})$, and $B_{\rm ref, x, y}(\tau = 10^{-4})$.
    \item Realistic LOS: Same as above, but we only incorporate the line-of-sight component of the chromospheric magnetogram, i.e. $\vec{B}_{\rm ref}(\tau = 10^{-1})$ and $B_{\rm ref, z}(\tau = 10^{-5})$.
\end{itemize}
All multi-height extrapolations use a potential field as side and top boundaries, as it leads to a general quality improvement. The 'Realistic' configurations aim to approximate recent observing capabilities, where only a limited number of magnetic field measurements are available (e.g., two $\tau$ surfaces). For the realistic configurations, we assume that the photospheric slice $\tau=10^{-1}$ is fixed at the bottom-boundary, while we map the chromospheric slice between 0 and 11.52 Mm ($z' = [0, 30]$ pixels), and set the initial estimated height $z$ to 1.15 Mm (3 pixels). For the configuration with split horizontal and vertical components, we use the same height configuration for the vertical component ($\tau=10^{-5}$). For the horizontal components ($\tau=10^{-4}$) we set the initial height estimate $z$ to 0.768 Mm (2 pixels) and the height range to $z' = [0, 7.68]$ Mm ($[0, 20]$ pixels).

We compare the resulting magnetic field extrapolations $\vec{B}_{\rm ext}$ to the ground-truth magnetic field from the MURaM simulation snapshot $\vec{B}_{\rm ref}$. Table \ref{table:muram} summarizes the results of our evaluation. Our height mapping model throughout outperforms the potential field approximation and the photospheric magnetic field extrapolations. For the single-height extrapolation the potential field boundary leads to a clear improvement, indicating that for a more complex magnetic field and a larger simulation volume, the potential field boundary is more favorable. For the individual configurations the performance is increasing with the inputs that are provided, where we achieve the best performance when using all available $\tau$ surfaces. We note that the realistic configuration with two layers largely outperforms single-height extrapolations when the full set of vector components is provided. When considering only the line-of-sight component of the chromospheric magnetic field $B_{\rm ref, z}(\tau = 10^{-5})$ the improvement is minor. The additional horizontal field (Realistic split) leads to a clear improvement over the baseline extrapolations, but provides a slightly lower performance than the configuration with the spatially aligned vector magnetic field (Realistic vector). The metrics for divergence-freeness ($L_{n,div}$) and force-freeness ($\sigma_J$) show that our extrapolations are consistent with the force-free model. Specifically, the divergence deviation is in the same range as for the potential field. We note that the MURaM reference is computed analogously, while the simulation is performed with a different stencil, which results in the larger divergence $L_{n,div}$.

\begin{figure*}
    \centering
    \includegraphics[width=\linewidth]{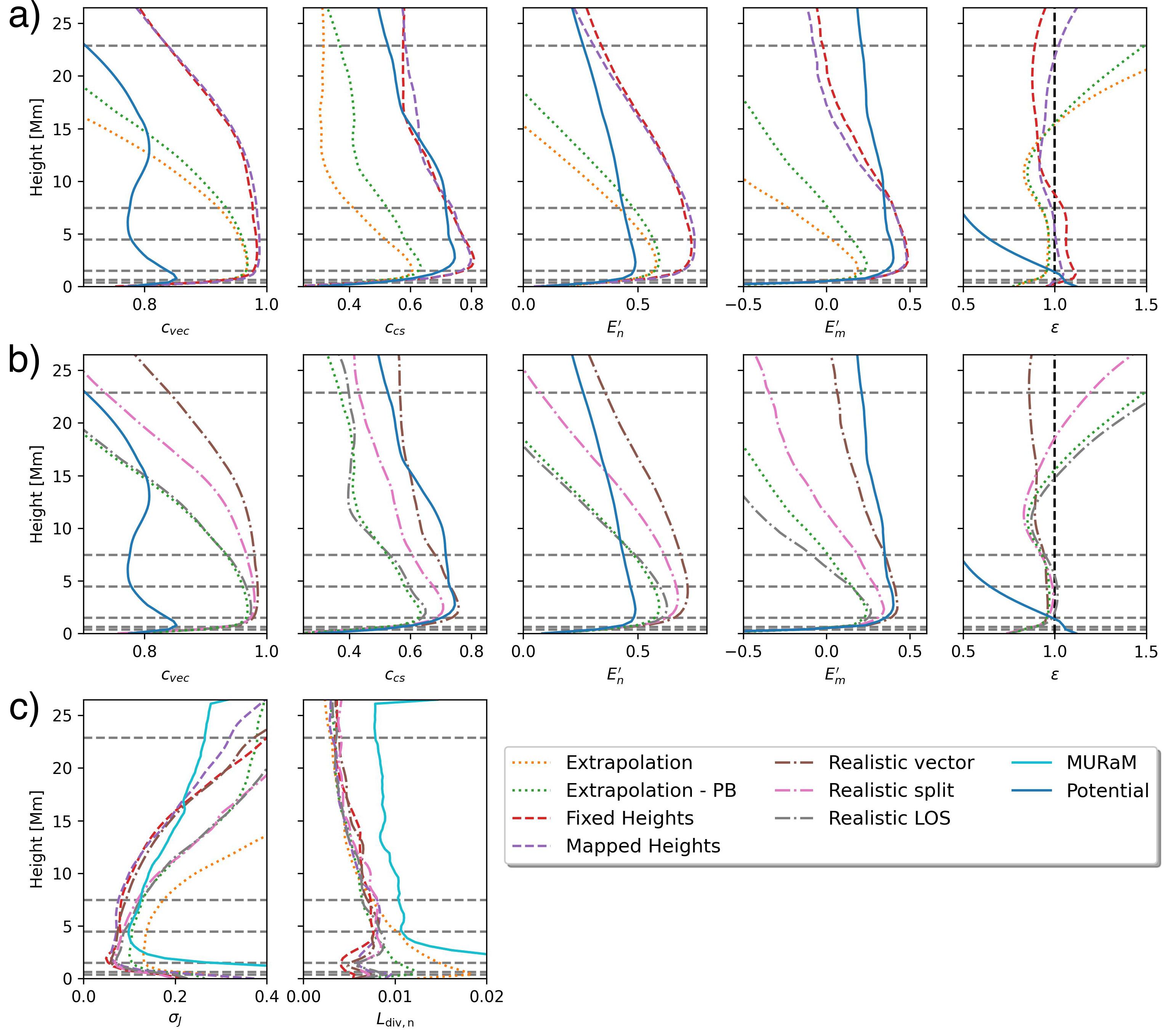}
    \caption{Evaluation of $\vec{B}_{\rm ext}$ as a function of geometrical height. The dotted horizontal lines indicate the average height of the $\tau$ surfaces. a) Comparison to $\vec{B}_{\rm ref}$ for the baseline single-height extrapolations (Potential, Extrapolation, and Extrapolation - PF) and the multi-height extrapolations (Fixed Height, Mapped Heights). b) Same comparison as in a), but for the realistic configurations (Realistic vector, Realistic split, and Realistic LOS). The best value of all performance metrics is 1. c) Evaluation of the force-freeness ($\sigma_J$) and divergence-freeness ($L_{\rm n, div}$) of the extrapolated and reference magnetic fields as function of geometrical height. The best value of both metrics is 0. We compare the reference magnetic field (MURaM), and our single-height (Extrapolation, and Extrapolation - PF) and multi-height extrapolations (Fixed Height, Mapped Heights, Realistic vector, Realistic split, and Realistic LOS). Note that the value range in all plots is adjusted to enhance the visibility of differences among the extrapolation methods.
    }
    \label{fig:height_metrics}
\end{figure*}

\begin{figure*}
    \centering
    \includegraphics[width=0.9\linewidth]{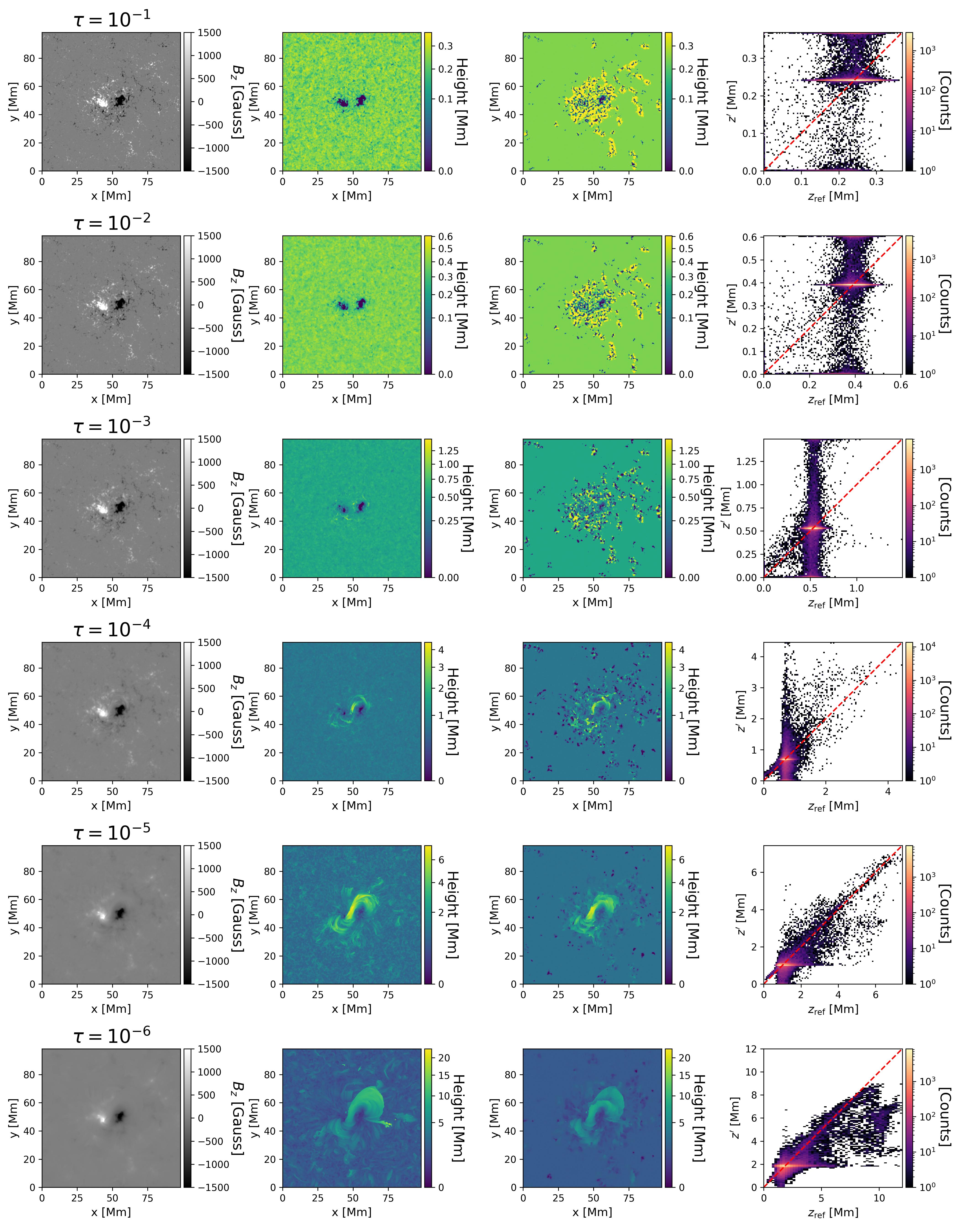}
    \caption{Comparison between the height surfaces at which magnetograms are formed at constant optical depths $\tau$ in the MURaM simulation $z_{\rm ref}(x, y, \tau)$ and the height surfaces at which they are inferred by our model $z'(x, y, z)$.
    The columns show $B_{{\rm ref}, z}(\tau)$, the ground-truth $z_{\rm ref}(x, y, \tau)$, the results of the height model $z'(x, y, z)$, and a 2D histogram of ground-truth and mapped heights. The red lines in the 2D histograms indicates the ideal one-to-one correlation. From top to bottom, the rows show decreasing optical depths $\tau$ (i.e., increasing geometrical heights).}
    \label{fig:tau}
\end{figure*}

To better understand the advantage that we obtain from the multi-height measurements we plot the metrics in dependence of height (Fig. \ref{fig:height_metrics}). The extrapolations from photospheric vector magnetograms quickly deviate from the reference field with increasing height. The use the multi-height measurements (fixed and mapped heights), show throughout an improvement in height and show even at greater heights ($>10$ Mm) a good agreement with the reference field. In particular, above 20 Mm the ratio of modeled and reference magnetic energy $\varepsilon$ is close to 1, indicating a more consistent approximation of the energy that is stored in the coronal magnetic field. Surprisingly, by only adding the $\tau=10^{-5}$ magnetic field, we achieve an almost equal performance. In contrast, the reduced line-of-sight information only leads to a small performance increase, particularly at lower heights ($\approx$ 2 Mm). The benefit from the full set of $\tau$ surfaces can be best seen from the relative total magnetic energy $\varepsilon$ low in the solar atmosphere ($\approx$ 1 Mm), where the additional magnetic field information prevents the initial underestimation of the photospheric magnetic field, which occurs due to the adjustment to the force-free condition in this layer (see $\sigma_J$ in Fig. \ref{fig:height_metrics}c). The overestimation of the magnetic energy at greater heights can be related to the increased uncertainties in the regime of weak magnetic fields. The comparison to the baseline potential field shows that this simple model is largely outperformed by the NLFF approach.

In Table \ref{table:height_muram} we summarize the results of the height differences. The use of the height mapping model leads to an overall improvement over the use of fixed heights. The comparison of the height differences  of the corrugated surfaces shows that the height mapping in the lower atmospheric layers does not improve over the simple assumption of the average height. We associate this with the stronger deviation from the force-free assumption in the lower atmospheric layers. In Fig. \ref{fig:height_metrics}c, the force-freeness and divergence-freeness is plotted as a function of height. This shows that the largest deviation from the force-free assumption occurs in the first few Mm of the reference magnetic field from MURaM and is similarly reflected by our extrapolations. This suggests that for photospheric heights the magnetic field data does not profit from a remapping.
The realistic configurations with a single $\tau$ layer show a large difference in height, which is related to the approximated initial height, which is computed from the ground-truth for the idealized configurations (Fixed Heights, Mapped Heights).

Fig. \ref{fig:tau} compares the inferred geometrical height maps associated with constant $\tau$ surfaces to the MURaM ground truth $z_{\rm ref}(x, y, \tau)$.
The model provides a realistic estimation of the height of formation of the magnetograms $\vec{B}_{\rm ref}(\tau)$, and their overall structure is well approximated. The best results are achieved at greater heights, where we expect low plasma density conditions (i.e., solar corona).
For $\tau \geq 10^{-3}$, sunspots are correctly mapped close to the bottom boundary of the simulation volume, but small-scale height fluctuations are not properly captured. This can be also seen from the 2D histograms, where quiet Sun regions are mostly mapped to constant heights and with increasing extent in height the distributions follow the ideal correlation (red line).

\begin{table*}[!h]
\centering
\caption{Quantitative comparison between $\vec{B}_{\rm ref}$ and $\vec{B}_{\rm ext}$.
(1) MURaM: Assessment of the force-freeness and divergence-freeness of the reference magnetic field. (2 - 4): Baseline (single-height) methods for comparison.
(5 - 6): comparison of mapped and fixed heights, where we consider all six magnetograms as input to our model. (7 - 9): realistic setting with a photospheric and chromospheric magnetogram ($\tau = \{10^{-1}, 10^{-5}\}$). For $B_{\rm ref}(\tau = 10^{-5})$ we consider a configuration with the full vector components and a configuration with the vertical component ($B_{\rm ref, z}$) only. For 8 we also consider the horizontal vector magnetic field of $B_{\rm ref}(\tau = 10^{-4})$ in addition to the vertical component $B_{\rm ref, z}(\tau = 10^{-5})$.  We compare the vector correlation ($C_{vec}$), angular difference ($C_{CS}$), the complement of the normalized vector error ($E'_n= 1 - E_n$), the complement of the mean vector error ($E'_m = 1 - E_m$), relative total magnetic energy ($\varepsilon$), the current-weighted average of the sine of the angle between the magnetic field and electric current density ($\sigma_J$), and the normalized divergence $L_{n,div}$.}
\label{table:muram}      
\centering                          
\begin{tabular}{| l || c c c c c c c |}        
\hline
 Method & $C_{vec}$ & $C_{cs}$ & $E'_n$ & $E'_m$ & $\varepsilon$ & $\sigma_J \times 10^2$ &  $ L_{\text{n,div}} \times 10^2$ [pixels$^{-1}$] \\    
\hline                        
  \hline
  1 MURaM & - & - & - & - & - & 60.09 & 3.66\\
  \hline
  \hline
  2 Potential & 0.80 & 0.64 & 0.40 & 0.27 & 0.89 & - & 0.89 \\
  \hline
  3 Extrapolation & 0.90 & 0.39 & 0.33 & -0.56 & 0.90 & 19.36 & 0.69 \\
  \hline
  4 Extrapolation - PB & 0.91 & 0.47 & 0.39 & -0.21 & 0.90 & 15.95 & 0.59 \\
  \hline
  \hline
  5 Fixed heights & 0.92 & 0.65 & 0.58 & 0.21 & 1.04 & 10.83 & 0.57 \\
  \hline
  6 Mapped heights  & \textbf{0.93} & \textbf{0.72} & \textbf{0.62} & \textbf{0.31} & \textbf{0.99} & 13.53 & 0.84 \\
  \hline
  \hline
  7 Realistic vector & 0.92 & 0.63 & 0.58 & 0.22 & 0.89 & 12.08 & 0.60 \\
  \hline
  8 Realistic split  & 0.91 & 0.55 & 0.50 & -0.02 & 0.89 & 13.02 & 0.61 \\
  \hline
  9 Realistic LOS  & 0.91 & 0.47 & 0.39 & -0.36 & 0.91 & 12.90 & 0.57 \\
\hline                                   
\end{tabular}
\end{table*}

\begin{table*}[!h]
\centering
\caption{Comparison of height maps from MURaM $\tau$ surfaces to estimated height maps. We compare the deviation from the reference height maps ($\Delta h$) in pixels and in percent for fixed heights, mapped heights, and in the realistic setting with two $\tau$ surfaces, where we consider configurations with the full vector components, split vertical and horizontal components, and only the line-of-sight component.}
\label{table:height_muram}      
\centering                          
\begin{tabular}{| l || c c |}        
\hline
 Method & $\Delta$h [Mm] & $\Delta$h [\%] \\    
\hline                        
  \hline
  5 Fixed heights & [0.03, 0.03, 0.04, 0.07, 0.30, 0.73] & [8.1, 5.2, 2.5, 1.6, 4.0, 3.2] \\
  \hline
  6 Mapped heights  & [0.04, 0.05, 0.08, 0.13, 0.24, 0.53] & [10.9, 8.1, 5.4, 2.8, 3.2, 2.3] \\
  \hline
  \hline
  7 Realistic vector & [-, -, -, -, 0.35, -] & [-, -, -, -, 4.7, -] \\
  \hline
  8 Realistic split & [-, -, -, 0.24, 0.41, -] & [-, -, -, 5.4, 5.5, -] \\
  \hline
  9 Realistic LOS  & [-, -, -, -, 0.58, -] & [-, -, -, -, 7.7, -] \\
\hline                                   
\end{tabular}
\end{table*}

\subsection{Application to SOLIS data}
\label{sec:observation}

Based on our previous evaluation using the MURaM simulation, we assume that incorporating multi-height information (e.g., slices from the chromosphere) increases the level of realism of our magnetic field extrapolations, even for fields that do not satisfy the force-free assumption.
Here we use the LOS magnetic field observations from SOLIS/VSM introduced in Sect. \ref{sec:data_observation}. As can be seen from our evaluation in Sect. \ref{sec:muram}, the height mapping at photospheric heights has limited application. Therefore we assume no vertical extend of the bottom boundary condition (i.e., photospheric vector magnetogram). For the internal condition from SOLIS (i.e., chromospheric line-of-sight magnetogram), we assume an initial formation height of $z = 2 $ Mm (5.56 pixels), according to the estimated formation height of Ca II 854.2 nm \citep{rodriguez2017radiative}. We set the height range of $z'$ to $[0, 10]$ Mm ([$0, 27.78$] pixels), similar to the height ranges computed from the MURaM $\tau$ surfaces. We normalize the magnetic field strength to 2500 Gauss and the spatial scaling to 320 pixels.

\begin{figure*}
    \centering
    \includegraphics[width=\linewidth]{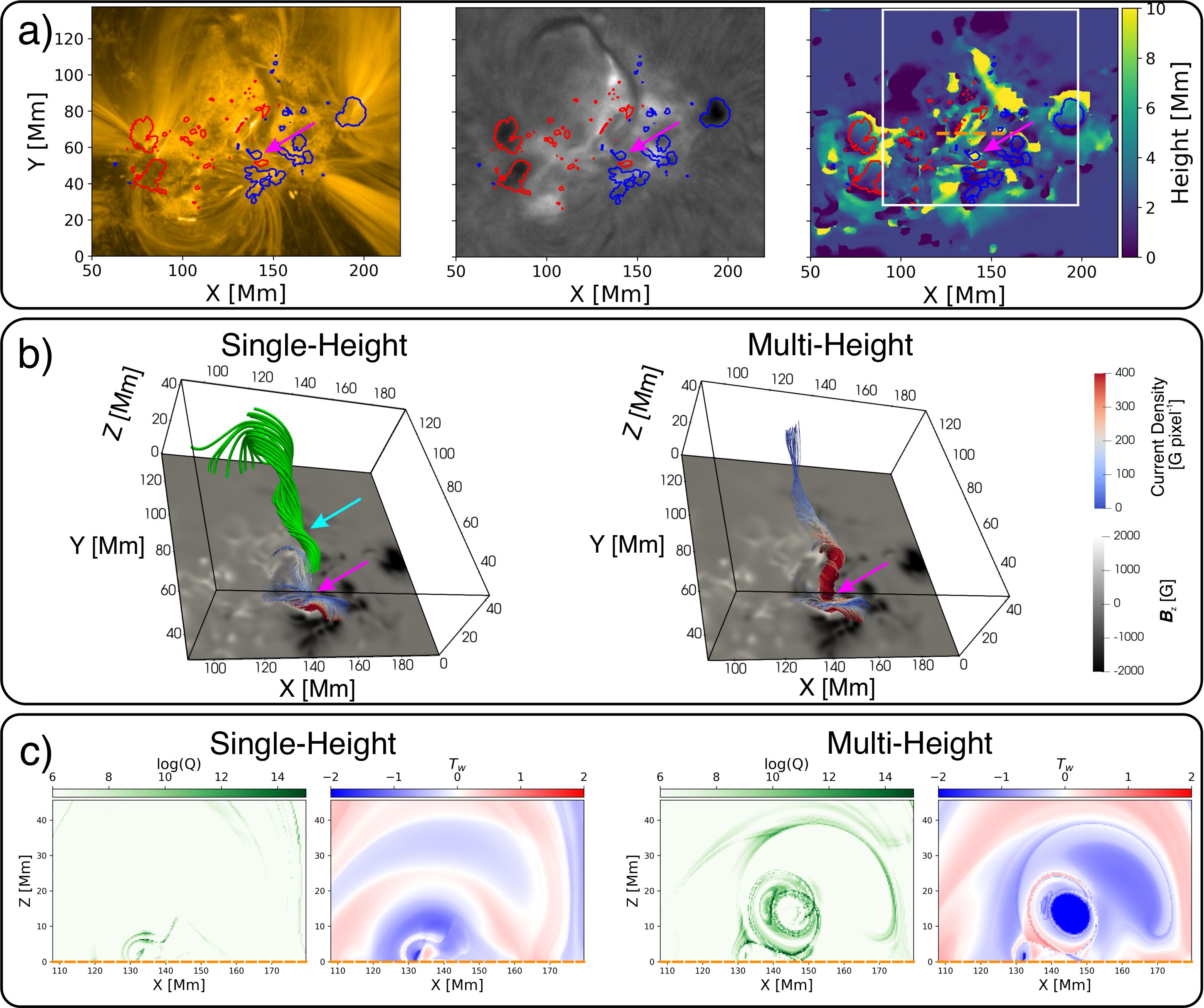}
    \caption{Application of our multi-height magnetic field extrapolation to observations from 2011-03-09 16:34. a) Reference EUV (left) and H$\alpha$ (middle) observation together with the estimated height from our height mapping model $z'$ (right) in the coordinate frame of the SHARP region (NOAA AR 11166; SHARP 401). The vertical magnetic field strength is overlaid as contours at -1000 and 1000 Gauss in blue and red, respectively. b) Comparison of magnetic field lines modeled with a single-height (photospheric HMI) extrapolation (left) and an extrapolation that includes also SOLIS/VSM chromospheric magnetic field information (right). The region corresponds to the white rectangle in panel a. The pink arrows indicate the position where we extract magnetic field lines. The blue arrow indicates the position where additional field lines were extracted for the single-height extrapolation. c) Squashing factor and twist number of the single-height (left) and multi-height (right) extrapolated magnetic field. The orange dashed line in panel a indicates the position where the slices are extracted.}
    \label{fig:observation}
\end{figure*}

In Fig. \ref{fig:observation} we show the result of our multi-height extrapolation in comparison to a photospheric magnetic field extrapolation. Panel a shows the corresponding EUV observation of the SDO/AIA 171~{\AA} filter \citep{lemen2012aia} and H$\alpha$ filtergram from the Kanzelhöhe Solar Observatory \citep[KSO; ][]{poetzi2021}, where we note the central elongated flux rope. From the extrapolations we extract field-line plots for comparison of the magnetic topology. Here, we extract field-lines at the same origin, indicated by the pink arrows (close to the supposed footpoint of the central filament). The comparison of the magnetic field topologies shows that the multi-height magnetic field measurement leads to a large difference in magnetic field configuration. For the multi-height extrapolation the magnetic flux shows a much higher twist (increased current density) and is strongly elongated. The single-height extrapolation recovers only a slightly twisted magnetic field and connects polarities already low in the solar atmosphere. We extract additional field lines, indicated by the blue arrow, to identify the flux rope, which does not connect to the strong polarities at the center of the active region. In panel c, we show a vertical slice of the squashing factor $Q$ and the twist number $T_\text{w}$ for both extrapolations. This additionally confirms that the multi-height extrapolation recovers a larger twist and a higher flux rope.

The estimated height map in panel a shows that the magnetic field close to the filament is at large heights ($>$10~Mm), which seems also indicated by the AIA observations. The strong magnetic field of the sunspots is mapped low in the solar atmosphere, in agreement with our results from Sect. \ref{sec:muram}. The height map shows artifacts north of strong magnetic polarities. These features are not present in our application to the MURaM simulation and are likely associated to the insufficient treatment of the line-of-sight component.

\section{Discussion}

In Fig. \ref{fig:height_metrics} we show the different quality metrics from Table \ref{table:muram} as function of height. Our height mapping module achieves for all heights the best agreement with the reference magnetic field data. The relative total magnetic energy $\varepsilon$ emphasizes that the multi-height magnetic field strongly improves the estimated energy distribution in height. In contrast, the single-height extrapolation and the potential field show strong deviations at higher atmospheric layers ($>5$~Mm). The additional magnetic field layers lead to only minor improvements (i.e., Multi Height vs Realistic), while the additional information of the horizontal components has a larger impact on the resulting magnetic field (comparison of Realistic vector, Realistic split, and Realistic LOS).

The evaluation of our method shows the importance of additional multi-height information for the coronal magnetic field extrapolations. Our experiments demonstrate that additional information provides for a more consistent magnetic energy distribution and a more realistic approximation of coronal magnetic fields, and that even one additional observational layer can drastically improve the outcome.

The comparison to the semi-analytical magnetic field solution shows that especially higher layers can improve the extrapolation (e.g., improvement of $E'_m$ from 0.71 to 0.98) and achieve results equivalent to the full boundary information. The height mapping model shows a consistent treatment of corrugation, where our method achieves an equivalent performance to geometrical slices. Furthermore, this allows us to estimate the corrugation of the surfaces, where errors are in the range of 1\%. The reduced information of only vertical magnetic field components leads to a quality decrease, but exceeds in every case the extrapolation from a single layer.

With the application to the MURaM snapshot we evaluate our method in a realistic setting of a non force-free field, and a realistic corrugation that is computed from the simulated radiation field. The extracted $\tau$ surfaces exhibiting a wide spectrum of corrugations, spanning from less than 1 Mm to exceeding 10 Mm, thus serving as a comprehensive and rigorous test scenario. The evaluation shows that chromospheric magnetic field measurements can more accurately capture the estimated magnetic field in the solar corona and match closer the energy distribution of the MURaM simulation (Fig. \ref{fig:height_metrics}). The comparison in Table \ref{table:muram} and Fig. \ref{fig:height_metrics}, shows a clear improvement over the baseline potential field and single-height extrapolation. The estimated heights show only a few percent in difference from the reference. Comparing the height estimates to the baseline (geometrical slices at the average heights) shows that the height mapping gives a better estimate for stronger corrugations that occur at greater heights. As can be seen from Fig. \ref{fig:tau}, our estimated $\tau$ surfaces also show a good spatial agreement with the reference corrugation map, particularly for large corrugations the topology is well captured (e.g., $\tau=10^{-6}$, $\tau=10^{-5}$). The MURaM snapshot presents a dynamic example of active flux emergence, introducing additional complexity for static magnetic field extrapolations that do not account for the temporal evolution of magnetic fields. In this context, the utilization of multi-height data can mitigate this shortcoming by integrating magnetic field information from the upper solar atmosphere.

The height mapping of the magnetic field is primarily driven by finding a better agreement of the magnetic field measurement to fit the force-free assumption. This is a consistent approach in settings where the force-free assumption is satisfied, but for non force-free conditions (i.e., photospheric and chromospheric heights) the correct mapping does not necessarily satisfy the force-free model. In other words, the height mapping tries to bring the observed magnetic field into a force-free state. This can be seen from Fig. \ref{fig:tau} where a proper estimate of the height surface is only achieved at chromospheric heights, while photospheric height estimates can not capture the small scale variations. For this reason, we consider photospheric observations as geometric slices, which prevents artificial height mappings that are driven by non force-free conditions. We use a height regularization that favors solutions close to the estimated line formation height (typically in the photosphere or chromosphere). With this we better constrain quiet Sun regions, where we assume that observations are obtained from lower layers in the solar atmosphere.

With the use of SOLIS data we demonstrate a first application of multi-height magnetic field extrapolations. The additional chromospheric observation provides a clear advancement. The comparison between the photospheric and the multi-height extrapolation in Fig. \ref{fig:observation} shows strong differences in magnetic field topology.

The distributions of $Q$ and $T_w$ in a vertical plane crossing the model magnetic flux rope recovered by the photospheric and multi-height NLFF modeling are shown in Fig. \ref{fig:observation}c. In agreement we find in both models a left-handed modeled magnetic flux rope, clearly separated from the untwisted field in its surrounding (as characterized by high values of $Q$). This separation is much more pronounced in the multi-height model, however. While the $T_w$ map pictures a low-lying, narrow, and elongated magnetic flux rope structure in the single-height extrapolation, it indicates the presence of a much more symmetric (in terms of cross-section shape) and elevated magnetic flux rope in the multi-height model. In addition, the multi-height extrapolation results in a more strongly twisted magnetic flux rope ($|T_w|$\,$\gtrsim$\,5.6) than the single-height model (peak value $|T_w|$\,$\gtrsim$\,2.0).

While the single-height extrapolation shows only a small flux rope, that quickly connects with the photospheric field, the multi-height extrapolation yields an elongated flux rope that is visually in good agreement with the EUV observation by SDO/AIA. While this first application suggests a better agreement with observational data, a further comparison of height estimates with stereoscopic measurement could further support the validity of our approach. A  shortcoming of our current approach is the insufficient correction for the line-of-sight observation. In other words, we approximate that the vertical coordinate aligns with the line-of-sight, which does not hold when we consider active regions that are more distant from the disk center. As a next step, the inclusion of the geometry (i.e., spherical coordinates), viewing perspective, and optically-thin plasma could improve the extrapolation.

Recent observing capabilities provide vector magnetograms from chromospheric heights (e.g., \cite{vissers2022sst_multi_height} using the Swedish Solar Telescope \citep{sst2003scharmer}). Our study suggests that the full vector information at layers above the photosphere is key to provide more realistic estimates of the coronal magnetic field and lead to better height estimations (c.f., Sect. \ref{sec:muram}). We further note that our method optimizes for divergence-free solutions, while this constrain can be also enforced by computing the vector potential. Within the PINN framework this a straight-forward extension. These extensions will be addressed in a future study.

A primary use case of magnetic field extrapolations is the study of energy build-up and flare related energy release processes. To this point, there is no readily available data set that provides multi-height observations at a sufficient cadence (of the order of 10 min) for the evolution during flares and/or coronal mass ejections, which currently limits the application of our method to single extrapolations in time. We emphasize that with this new method, we provide a consistent approach to incorporate magnetic field measurements from multiple heights into a single magnetic field extrapolation, which paves the way for future high-cadence observing series for a better understanding of the energy build-up in active regions, the magnetic reconfiguration and energy release in flares and associated coronal mass ejections.

\section{Data availability}
All our extrapolation results are publicly available.

Our codes are publicly available. We provide Python notebooks that perform extrapolations for arbitrary regions without any pre-requirements. See project page: \url{https://github.com/RobertJaro/NF2}.

The SDO HMI and AIA data is provided by JSOC (\url{http://jsoc.stanford.edu/}). We provide automatic download scripts that use SunPy \citep{sunpysoftware2020, glogowski2019drms}.

\section{Acknowledgments}
This research has received financial support from the European Union’s Horizon 2020 research and innovation program under grant agreement No. 824135 (SOLARNET). JT and AV acknowledge the Austrian Science Fund (FWF): P31413-N27.
The authors acknowledge the use of the Skoltech Zhores cluster for obtaining the results presented in this paper \citep{zacharov2019zhores}. This work utilizes SOLIS data obtained by the NSO Integrated Synoptic Program (NISP), managed by the National Solar Observatory, which is operated by the Association of Universities for Research in Astronomy (AURA), Inc. under a cooperative agreement with the National Science Foundation.
This research has made use of SunPy v3.0.0 \citep{sunpysoftware2020, sunpycommunity2020}, AstroPy \citep{2013A&A...558A..33A}, PyTorch \citep{pytorch2019_9015} and Paraview \citep{ahrens2005paraview}.

%

\vspace{5mm}
\facilities{SDO (HMI, AIA), SOLIS (VSM).}


\software{Astropy \citep{2013A&A...558A..33A,2018AJ....156..123A},
          Sunpy \citep{sunpycommunity2020, sunpysoftware2020, glogowski2019drms},
          Pytorch \citep{pytorch2019_9015},
          Paraview \citep{ahrens2005paraview}.
          }






\bibliography{sample631}{}
\bibliographystyle{aasjournal}



\end{document}